\documentclass[final,5p,times]{elsarticle}
\usepackage{graphicx}
\usepackage{amsmath}
\usepackage{amssymb}
\usepackage{natbib}
\usepackage{epsfig}
\usepackage{subfigure}
\usepackage{enumerate}
\usepackage{hyphenat}
\biboptions{comma,square}

\def\araa{\rm{ARA\&A}}             
\def\apj{\rm{ApJ}}                 
\def\apjl{\rm{ApJ}}                
\def\apss{\rm{Ap\&SS}}             
\def\aap{\rm{A\&A}}                
\def\mnras{\rm{MNRAS}}             
\def\prd{\rm{Phys.~Rev.~D}}        
\def\pasp{\rm{PASP}}               
\def\nat{\rm{Nature}}              

\journal{Astroparticle Physics}

\begin{document}

\begin{frontmatter}

\title{Uncovering neutrinos from cosmic ray factories: the Multi Point Source method}

\author[label1]{Yolanda Sestayo}
\ead{Yolanda.Sestayo@ub.edu}
\author[label2]{Elisa Resconi}
\address[label1]{Departament d'Astronomia i Meteorologia, Universitat de Barcelona (IEEC-UB), Mart\'{\i} i Franqu\`es 1, E-08028, Barcelona, Spain}
\address[label2]{T. U. Munich, D-85748 Garching, Germany}

\begin{abstract}

We present a novel method for the search of high energy extraterrestrial neutrinos in extended regions. The method is based on the study of the spatial correlations between the events recorded by neutrino telescopes. Extended regions radiating neutrinos may exist in the Galaxy due to the hierarchical clustering of massive stars, the progenitors of all the Galactic accelerators known so far. The neutrino emission associated to such extended regions might be faint and complex due to both the escape of cosmic rays and the intricate distribution of gas in the environment of the accelerators. We have simulated extended neutrino emission over an area of $10^{\circ} \times 10^{\circ}$, where the intensity fluctuations across the region are modelled as a Gaussian random field with a given correlation structure. We tested our proposed method over realizations of this intensity field plus a uniform random field representative of the spatial distribution of the atmospheric neutrino background. Our results indicate that the method proposed here can detect significant event patterns that would be missed by standard search methods, mostly focused in the detection of individual hot spots.

\end{abstract}


\end{frontmatter}

\section{Introduction}
\label{s:1}

The search for the sources of cosmic rays is one of the most standing issues in high-energy astrophysics. Major advances in the field during the last years have been realized by the considerable number of instruments dedicated to Galactic cosmic ray studies \cite{Fermi09_LAT, Hess, Magic, Veritas, Milagro, Pamela}. The on-going measurements of the cosmic radiation from the Galaxy \cite{FermiDiffuse09, Fermi09_epm, PamelaAnomaly} are helping to construct a comprehensive picture about the sources of cosmic rays, the mechanisms of propagation, and the interaction of cosmic rays with the gas and radiation fields of the Galaxy \cite{Strong07}. Currently, gamma-ray astronomy has a leading role in the exploration of the sky at the very high energies. Measurements of the spatial and spectral distributions of the diffuse Galactic gamma-rays provide information about the propagation and interaction of Galactic cosmic rays \cite{Fermi2010_perseus, Fermi2012_cygnus, FermiDiffuse2012}, whereas localized excesses of gamma-ray emission with respect to models of cosmic ray propagation can be interpreted as the result of cosmic ray interactions in regions of enhanced matter density (which are below the spatial resolution of current cosmic ray propagation models), or due to the presence of Galactic accelerators injecting high energy particles which interact close to their sources of origin \cite{HESS2012_Wd1, MGROCygnus, Veritas09_SNR, Fermi2011_cygnusbubb}. The GeV gamma-ray diffuse emission is being measured by the Fermi satellite \cite{FermiDiffuse09}, showing that gamma-rays are produced throughout the Galactic disk by the interactions of cosmic rays with the gas and radiation from the interstellar medium, after their diffusion in the Galactic magnetic fields. The Galactic TeV diffuse emission is within reach of ground Cherenkov telescopes like H.E.S.S. \cite{HESSDiffuse06} and Milagro \cite{MilagroDiffuse08}, instruments which also have provided the first evidence of particle acceleration up to TeV energies from a number of individual sources \cite{TeVCat}.

Further progress is expected in the upcoming future with neutrino detectors such as IceCube \cite{IceCube09}. In the energy range 300 GeV $<$ E $<$ 10 PeV, the IceCube experiment at the South Pole uses the Earth as a filter to observe neutrinos from the whole northern sky. Its large field of view and dynamic range offer a unique opportunity to unveil the sites of both production and interaction of the high energy cosmic rays from the Galaxy. As in the case of hadronic gamma-rays, neutrino emission implies the efficient acceleration of cosmic rays in Galactic sources, as well as their interaction with matter and radiation. Surveys of atomic and molecular gas \cite{HI_LAB,Dame01}, combined with infrared and optical data, have provided large-scale maps of the Milky Way's stellar and gas distribution with unprecedented spatial resolution \cite{Churchwell09MW}, despite the uncertainties in the distance estimations. From these observations we know that the progenitors of all the Galactic accelerators known so far, massive stars, are not located at random within the Galactic disk, but rather showing a hierarchical structure. 

All present-day star formation appears to take place in giant molecular clouds. Massive open star clusters do not form in isolation, but tend to be clustered themselves in the so-called clusters complexes \cite{Elmegreen2011}, associated with giant molecular clouds following the spiral arms of the Milky Way disk \cite{DeLaFuenteMarcos08, Schneider06, Efremov78}. De la Fuente Marcos \& De la Fuente Marcos \cite{DeLaFuenteMarcos08} found statistical evidence of the existence of at least five dynamical families of young and massive open clusters in the solar neighborhood (distance $<$ 2.5 kpc), which may extend up to 10 degrees in Galactic longitude. The relevant scenario for high energy astronomy is one in which we can find groups of stellar open clusters in an evolutionary stage in which the combined winds of massive stars and supernova explosions dominate the energetics of the region. This establishes the existence of potential Galactic cosmic ray factories in regions of massive star formation \cite{Higdon} which may well extend several degrees in the sky, possibly affecting the spatial distribution of neutrino events recorded with neutrino telescopes.

In this paper we focus on the potential of neutrino telescopes such as IceCube to detect a significant neutrino event pattern associated with these active regions of the Galaxy. This could point to the location and distribution of Galactic cosmic ray and neutrino sources during the first years of the IceCube experiment, when sensitivity might be still too limited for the detection of individual steady point-sources. 

A possible neutrino emission in these regions may show an intricate, and likely faint, intensity pattern, due to the complex structure of the matter in regions harboring massive stars. On one hand we have the clustering of potential accelerators, where high energy cosmic rays are originated, and on the other hand the available target for cosmic ray interactions, consisting of the gas left from the parent molecular cloud, which has not been used for star formation nor evacuated from the region by the radiation and winds from the stars and supernova explosions. 

Both observations and theory about the spatial structure of the molecular and atomic gas in a large range of environments have shown that it presents some degree of spatial correlation, usually characterized in terms of the power spectra \cite{Stutzki98, Falgarone04}. The causes of the observed spatial correlations are manifold, but large scale turbulence and some local effects like the radiation pressure and winds from the massive stars and supernova explosions are the most important \cite{ElmegreenScalo2004, Schneider2011, Khalil2004, Lazarian2000}. These effects produce spatial fluctuations in the gas density, which in turn, may contribute to the fluctuations of a possible neutrino signal inside an area populated with cosmic ray sources. 

We offer here an alternative to the search of these potential Galactic cosmic ray factories, in a situation in which little knowledge about the high energy processes taking place is available. We introduce here the use of analysis of the spatial correlations between neutrino events for the discovery of Galactic cosmic ray sites. We will study the advantages and potential of correlation analysis with respect to standard searches for neutrino sources. The method presented here and the standard search methods are tested on simulated event patterns, in which the physical origin of the possible neutrino signal is translated into statistical properties of the neutrino intensity field. 

Modelling the exact characteristics of a possible neutrino intensity field in a particular region of the Galaxy is out of the scope of this paper, and the examples considered aim to illustrate the performance of search methods. Within the region under study, we assume that the fluctuations in the intensity of the neutrino signal vary spatially according to a Gaussian random field with a given correlation structure. The last source of neutrinos we consider is the isotropic foreground of atmospheric neutrinos, the statistics of which is consistent with a uniform Poisson distribution. 

The paper is organized as follows. In section 2 we describe the simulations of the event pattern. In section 3 we present our proposed method for the analysis of extended regions. Section 4 shows the performance of both our correlation method and standard search methods from the results obtained with the simulated event patterns. Section 4 has the discussion. 

\section{Simulations of the event pattern}
\label{s:2}

An event pattern consists of a set of event locations $\bf{r} = (r_i, r_j,..., r_N)$ within a region (R), at which N events have been recorded. In the simplest case, the data set comprises only the event locations. However, in some cases we may have additional information related to the events which might have a bearing with the nature of the analysis. For instance, the arrival time of the events, for analysis of variable sources \cite{JBTime}, or the energy of the events for a higher signal to noise discrimination power \cite{JBunbinned}. These cases correspond to what is known as marked point pattern. Although this information can be incorporated in any analysis, in this paper we have not considered marks in the event pattern, and we focused only on the spatial distribution of events.

\subsection{Simulations of the astrophysical neutrino signal}
\label{ss:2.1}

A spatial event pattern can be thought of as the realization of a spatial process. In many astrophysical scenarios, such physical processes depend on random components, and they are usually modelled in terms of random fields \cite{MartinezSaar2001}. 

We adopted a Gaussian random field with a given correlation structure for the intensity fluctuations of a possible neutrino signal inside a region that is much larger than the angular resolution of the experiment. 

If the field is Gaussian, we can determine it completely given its mean and correlation function (or power spectrum), which facilitates the simulations. The starting point of the simulations of Gaussian random fields is the convolution of the power spectrum of the field with a white noise (i.e., correlation free) random signal \cite{Salmon96}:

\begin{equation}
I(\boldsymbol {r}) = \int d^{3}k e^{i\boldsymbol{k}\cdot \boldsymbol{r}}\sqrt{P(\boldsymbol{k})}W(\boldsymbol{k}) 
\label{eq:grf}
\end{equation} 
which gives the intensity of the field at each point $\boldsymbol {r}$ from the power spectrum $P(\boldsymbol{k})$, related to the spatial correlation function $C(\boldsymbol{r})$ by the Fourier transform: $P(\boldsymbol{k}) = F^{-1}C(\boldsymbol{r})$. $W(\boldsymbol{k})$ is the white noise, which gives the random amplitude and phase to the field. 

If the angular extent of R is large, a spherical harmonic decomposition of the field would be needed. In this case, the field would have the spectral representation:

\begin{equation}
I(\boldsymbol {r}) = \sum_{l=0}^{\infty}\sum_{m=-l}^{l}a_{l,m}Y_{l,m}(\boldsymbol{r})
\label{eq:spgrf}
\end{equation}
where ${Y_{l,m}}$ is a basis of spherical harmonics, and the coefficients ${a_{l,m}}$ are complex uncorrelated random variables that depend only on the covariance or power spectrum of the field $P_{l}$.

In our examples, R is small enough as to assume Euclidean space. We generated the Gaussian random fields through Fourier phase randomization by discretizing equation \ref{eq:grf} with a Cartesian mesh defined over R. For simplicity, we assumed that R is a square region of length L = $10^{\circ}$, and used a mesh of 512 $\times$ 512 nodes, giving a 0.02$^{\circ}$ sampling interval of R, much smaller than the angular resolution of neutrino telescopes. 

The sampling of R is done in Fourier space directly assuming a power spectrum for the intensity field. The Fourier domain is commonly used in the simulations of random fields because the random variables at different points are statistically independent, whereas in the spatial domain they have long-range correlations that are difficult to sample. 

For simplicity, we assumed an isotropic, homogeneous random field; that is, the correlation between two points $\boldsymbol{r_1}$ and $\boldsymbol{r_2}$ depends on the length $\tau = |\boldsymbol{\tau}|$ of the vector $\boldsymbol{\tau} = \boldsymbol{r_1} - \boldsymbol{r_2}$, but not on its direction. We adopted a power-law power spectrum $P(|\boldsymbol{k}|) = (\alpha + |\boldsymbol{k}|^2)^{-\beta }$, which has an associated covariance function described also by a power-law plus an exponential decay, with $\alpha, \beta$ as the parameters that regulate the decay rate of the long range correlations with distance: $C(\tau) \propto \tau^{\beta -1}e^{-\alpha\tau}$. 

\begin{figure}
\centering
\includegraphics[clip,width=0.45\textwidth, keepaspectratio]{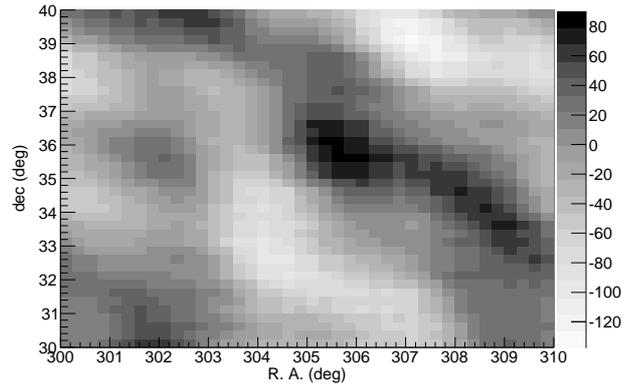}
\caption{Realization of a Gaussian random field with power spectrum $P(|\boldsymbol{k}|) = (1 + |\boldsymbol{k}|^{2})^{-2}$ over the $10^{\circ} \times 10^{\circ}$ region used in our examples. In this specific realization units are arbitrary, the intensity of the field is normalized with respect to a reference value. In the following examples the signal will be normalized to a given number of events.}
\label{fig:1}
\end{figure}

Figure \ref{fig:1} shows a realization of an isotropic, homogeneous Gaussian random field with the power spectrum described above with $\alpha = 1, \beta = 2$. In figure \ref{fig:1} the color scale has arbitrary units, but the obtained spatial distribution was used for the representation of the astrophysical neutrino signal, where the intensity will be normalized to a certain number of signal events in the final data sample of neutrino telescopes. The test region over which we inject the signal is defined in the range in equatorial coordinates $30^{\circ} < $dec$ < 40^{\circ}, 300^{\circ} < $R.A.$ < 310^{\circ}$. 

The instrumentation introduces an additional correlation in the intensity pattern due to the errors in the reconstruction of the event directions. Therefore, for the analysis of the event pattern in section we convolved the signal field of figure \ref{fig:1} with the Point Spread Function (PSF) typical of neutrino telescopes. The effect of the PSF is to smear the signal over some area, according to its shape. We assume the PSF has a Gaussian profile with $\sigma = 1$, which corresponds to a median of $0.7^{\circ}$, similar to what is expected for the IceCube neutrino detector \cite{IceCube40}. 

\subsection{Simulation of the background of atmospheric neutrinos}
\label{ss:2.2}

A simulated sample consisting only of atmospheric neutrino events was constructed injecting uniform random events on a sphere. In reality, atmospheric neutrinos reach the detector with a zenith dependent component \cite{IceCubeAtmo}, but within a declination band of $\sim 10^{\circ}$ scales like in the examples we consider here, the spatial distribution of atmospheric neutrino events in the final sample is uniform random, where the intensity at each point follows a Poisson process.

A total of 50000 background events were injected randomly through the northern sky, a number which could be representative of the total atmospheric neutrino events with E $> 500$ GeV expected for the IceCube detector in its final configuration. This produces and expectation of 207 background events in the $10^{\circ} \times 10^{\circ}$ region within the declination band $30^{\circ} < \delta < 40^{\circ}$ that we have considered in our examples.

\section{Analysis of the spatial correlations: the Multi Point Source method (MPS)}
\label{s:3}

Methods to account for the spatial correlations in the data are extensively used in astronomy \cite{AdayMergers}, starting from the work of Peebles and co-workers to study the large-scale matter distribution in the Universe through the two-point correlation function of galaxies \cite{Peebles}. In high energy astrophysics, the searches for correlations in the data have been applied mainly to detect anisotropies in the spatial distribution of cosmic-rays \cite{Westerhoff04, Chad04} and neutrinos \cite{abbasi09AMANDA}. 

The most commonly used estimators for the two-point correlation function in astronomy are based on pair counting (see Kerscher et al. (2000) \cite{Kerscher2000} for a comparison of different estimators). Following the notation of Szapudi \& Szalay (1998) \cite{SzapudiSzalay98}, the number of event pairs within a distance $\it{r}$ is defined as 

\begin{equation}
\rm{P_{DD}(r)} = \sum_{x \in D}\sum_{y \in D}\Psi_{r}(x,y)~~~~(x \neq y) 
\end{equation}
\begin{equation}
\rm{P_{RR}(r)} = \sum_{x \in R}\sum_{y \in R}\Psi_{r}(x,y)~~~~(x \neq y)
\end{equation} 
where the summations run over event coordinates in the data sample D, and in a sample R of randomly distributed events. $\Psi_{r}(x,y) = 1$ only if a certain pair selection criteria are satisfied in the real (random) sample, and is equal to 0 otherwise. After the normalization of the number of pair counts:

\begin{equation}
\rm{DD(r)} = \sum{\rm{P_{DD}(r)}}{\rm{N(N-1)}}
\end{equation}
\begin{equation}
\rm{RR(r)} = \sum{\rm{P_{RR}(r)}}{\rm{N_{R}(N_{R}-1)}}
\end{equation}
with N and N$_R$ being the total number of events in the real and random sample, the so-call natural estimator of the two-point correlation function, $\eta_n$, is:

\begin{equation}
\eta_{n}(r) = \frac{\rm{DD}}{\rm{RR}} - 1
\end{equation}
which represents the excess probability, with respect to a random distribution, of finding an event at a distance $\it{r}$ of another event. 

In this section, we introduce this formalism for the detection of high energy neutrinos from an extended region. The classical two-point correlation function method, briefly described above, was adapted to our specific case and optimized for discovery. The aim of our method is to determine whether the observed spatial properties of neutrino events inside the area under investigation are compatible with those expected from a distribution of background-only events (atmospheric neutrino events), or if an extraterrestrial component has to be invoked in order to explain the observations, and at which level the background-only hypothesis is rejected. 

Due to the differences our approach poses to the natural definition of the two-point correlation function, we defined our test statistics as a $\emph{clustering function}$, and in what follows we will refer to our proposed method as the $\emph{Multi Point Source}$ (MPS). Here we describe how the MPS method is defined. 

Let R be the region under study, in which a total number of $N_{inside}$ events have been registered in the final data sample at the locations ($r_1, r_2,...r_{N_{inside}}$), defined by the coordinates $r_i = (\theta_i, \phi_i)$ on the sky. MPS makes a two-point sampling of R with circular bins of variable area A = $\pi \Theta^2$ centered at the locations of each of the events inside the region, $r_i$. In the sampling of the region, the angular distance, $\Theta_{ij}$, from an event $\emph{i}$ located inside R to an event $\emph{j}$ (at any location) is measured for each of the $N_{inside}$ events present in R ($\emph{i}$ = $1,...,N_{inside}$; $\emph{j}$ = $1,...,N_{total}$). The number of pairs $\emph{ij}$ is measured as a function of the angular separation and the histogram of event pairs as a function of the angular distance $\Theta$ is constructed from these measurements. 

The analysis then makes use of a scale dependent cumulative clustering function, $\Phi(\Theta)$, defined as the excess, with respect to the null hypothesis, in the number of event pairs within a certain angular distance $\Theta$:

\begin{equation}
\Phi(\Theta) = \frac{\int_0^{\Theta}DD(\Theta) d\Theta}{\int_0^{\Theta}RR(\Theta) d\Theta}
\label{eq:clusfunc}
\end{equation}
where $DD(\Theta) = \displaystyle\sum_{ij}DD_{ij}$, $RR(\Theta) = \displaystyle\sum_{ij}RR_{ij}$, and the sum runs over all non-repeated pairs in the real data sample and in the background case, respectively.  In our case, $DD_{ij} (RR_{ij}) = 1$ only if either the event $\it{i}$ or the event $\it{j}$, or both, are within the region under study, and it is equal to zero otherwise. That is, MPS does not correlate events with specific locations, but rather it considers each event inside the region as a point source in order to determine its degree of correlation with the rest of the events. This feature is precisely what names the method as "Multi Point Source". With this definition we measure both the intensity of the process that generated the observed neutrino event pattern as well as its correlation structure.

The ultimate result of the analysis in neutrino telescopes is the p-value of the observation, that is, the probability that the observed event pattern is just a realization of the atmospheric neutrino background field. A discovery is claimed when this probability is below $2.8 \times 10^{-7}$, which corresponds to a 5$\sigma$ detection. To obtain the p-value of an observation, we must therefore know the distribution of our test statistics under the background-only hypothesis. In the MPS method, the test statistics is a function of the angular scale $\Theta$. For each $\Theta$ examined, we must construct the probability distribution of $\Phi(\Theta)$. The number of background events in a given area $A = \pi \Theta^2$ is produced by a Poisson process; as a consequence, the number of pairs of background events within a distance $\Theta$ follows a Gamma distribution. However, with the definition of event pairs in MPS, the parameters of the Gamma distribution are difficult to estimate from observables, and we obtained the distributions with Monte Carlo simulations.

\section{Analysis of the event pattern}
\label{s:4}
In this section, we present our results about the ability of both a stardard search method and the MPS method to find a significant neutrino events pattern over extended regions. The tests were performed over simulated skymaps, in which a region R of $10^{\circ} \times 10^{\circ}$ exhibits neutrino emission with some correlation structure (section \ref{ss:2.1}).

Standard search methods deal with the identification of hot spots, defined as a spatial concentration of events. The search for hot spots is done by mapping the sky using different techniques; the ones commonly used in neutrino astronomy are based on either "classical" $\emph{binned}$ methods, or likelihood-based methods with different degrees of sophistication in the modelling of the data \cite{JBunbinned, JuananEM}. Regardless of the method used in the analysis, the searches have been usually optimized for the discovery of point-like emission, and hence correlations of events beyond the single point-source scale do not enter in the analysis. The exploration of the sky, seeking for sources of high energy neutrinos, is done by superimposing a grid over the event distribution. Classical binned methods count the number of events within a circular fixed-area bin centered at each of the nodes in the grid, resulting in estimates of the intensity of the processes that generate the observed event pattern at each sample point. The grid step is much smaller than the angular resolution achieved in the analysis, and the bin size is usually optimized to have the best signal to noise ratio for point-sources. Likelihood methods use the same type of scan analysis, but they give a probabilistic assignment of an event to a component (signal or background) according to its distance to the point at which the intensity is being estimated. At each grid point, the likelihood of a mixture model of signal plus background is compared to a pure background hypothesis, and the region under study is imaged as a probability density map. 

The performance of both classical and likelihood methods are similar in terms of discovery, as long as they use the same event information \cite{JBunbinned}. Given that the event density maps obtained from a binned analysis offer a clear visual inspection of the spatial event pattern, as well as a more straightforward comparison with the results from MPS, we used a classical binned method as the standard scan analysis of our simulated skymaps. 

The scan of the test region of $10^{\circ} \times 10^{\circ}$ was done in steps of $0.25^{\circ} \times 0.25^{\circ}$. At each point, the local event density was measured and compared to the one expected under the background hypothesis, where by local we mean within a fixed-area bin around the sample points. 

Both the MPS and the binned scan of R were applied to background+signal samples and the corresponding significances were calculated with the p-value of the outcomes of each method. The probability distributions were computed after the analysis of $\it{n}$ random background-only samples with both the MPS method and with a binned scan of R. Each of these simulated data sets yields $\it{n}$ event patterns with a different number $\it{N^{(n)}_{inside}}$ of background events inside R. The average clustering function in the background case and its dispersion at different angular scales is obtained after applying MPS over $\it{n}$ = $10^4$ background-only data sets. The obtained probability distributions were fitted to a Gamma distribution. The binned scan was applied to the same background-only datasets and the distribution of the number of events within the search bin for each of the nodes in the grid was fitted to a Poisson distribution. 

In the next sections, we present our results under two different scenarios: a situation of high signal to noise ratio (S/N), and a situation in which the S/N is very low. The case of a high S/N underlines the differences in the information extracted from the same event pattern with the two different methods used in this paper. The study of a low-signal case illustrates how the MPS and a standard scan of R deal with the background dominated samples characteristic of neutrino telescopes.  

\subsection{High S/N case}
\label{ss:5.1}

In this example a total of 100 signal events were injected randomly following the distribution of figure \ref{fig:1} convolved with a Gaussian with $\sigma = 1$, representative of the PSF of current neutrino telescopes. On top of the signal, background events were randomly generated as explained in section \ref{ss:2.2}. 

\begin{figure*}[!ht]
\centering
\subfigure[Significance map of event densities at $0.5^{\circ}$ scales]{\label{fig:2a}\includegraphics[width=0.45\textwidth,keepaspectratio]{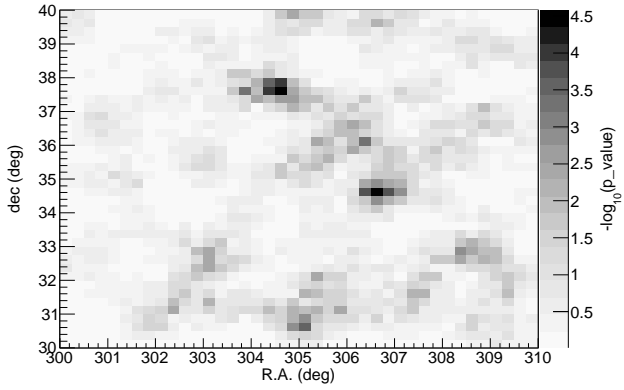}} \hspace{20pt} 
\subfigure[Significance map of event densities at $1^{\circ}$ scales]{\label{fig:2b}\includegraphics[width=0.45\textwidth,keepaspectratio]{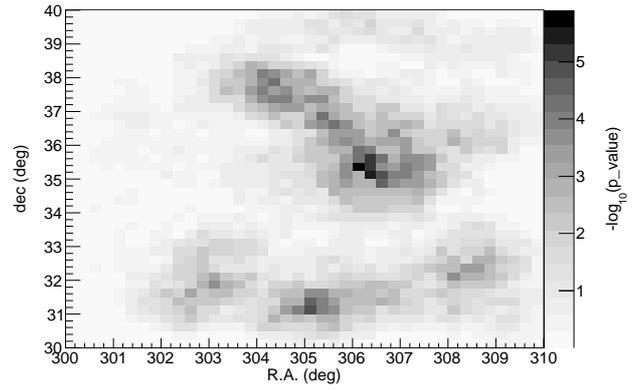}} \hfill \\ 
\subfigure[Significance map of event densities at $1.5^{\circ}$ scales]{\label{fig:2c}\includegraphics[width=0.45\textwidth,keepaspectratio]{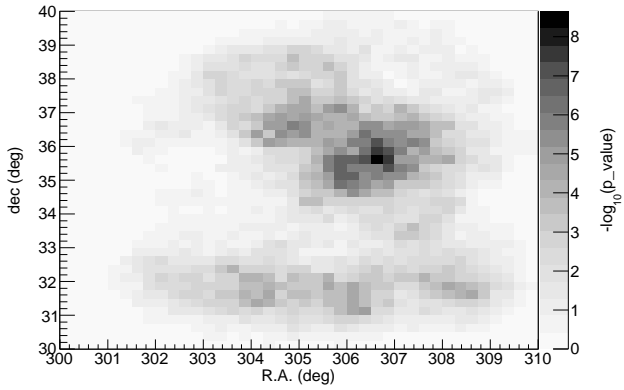}} \hspace{20pt}
\subfigure[Significance map of event densities at $2^{\circ}$ scales]{\label{fig:2d}\includegraphics[width=0.45\textwidth,keepaspectratio]{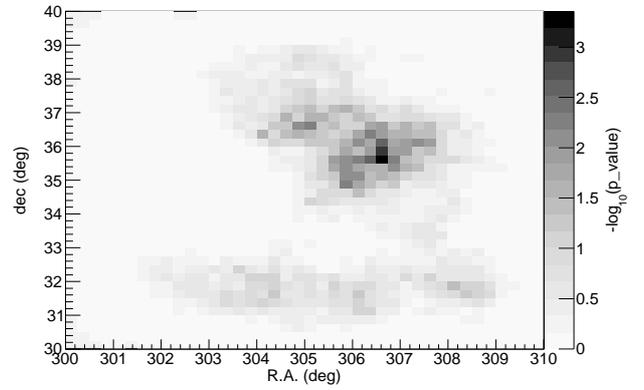}} \hfill \\ 
\caption{Results from the binned scan. The color scale is in units of standard deviation.}
\label{fig:2}
\end{figure*}

Fig.~\ref{fig:2} shows a realization of the signal+background field imaged as a significance event density map from the results of the binned scan of R with circular bins of radius $0.5^{\circ}$, $1^{\circ}$, $1.5^{\circ}$, $2^{\circ}$. In this simulated data set, a total of $N_{inside} = 294$ signal+background events have fallen within R, yielding a 6$\sigma$ excess of events inside the entire R, clearly representing a high S/N case. 

\begin{figure*}[!ht]
\centering
\subfigure[Clustering function]{\label{fig:3a}\includegraphics[width=0.45\textwidth,keepaspectratio]{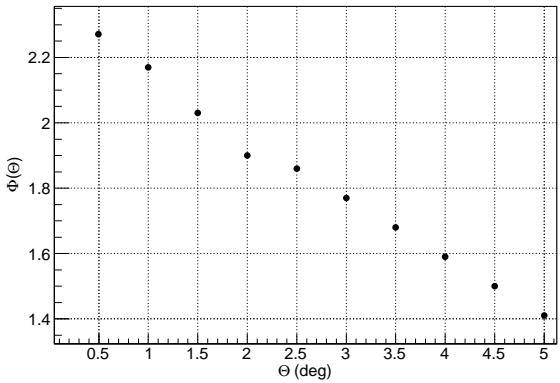}} \hspace{20pt}
\subfigure[Significance of the clustering function at each of the angular scales tested.]{\label{fig:3b}\includegraphics[width=0.45\textwidth,keepaspectratio]{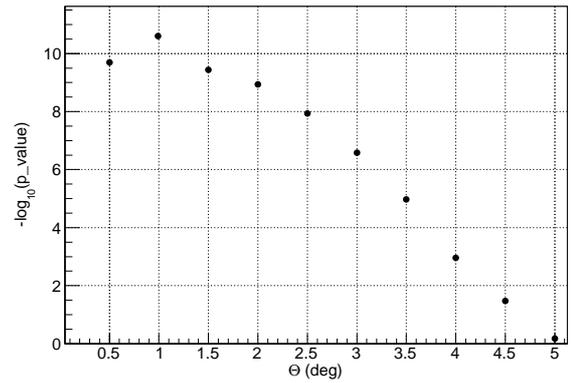}} \hfill \\ 
\caption{Results from the MPS}
\label{fig:3}
\end{figure*}

The appearance of the resulting images is a combination of diverse effects: the spatial structure of the signal intensity field, the smearing of this signal because of the instrumental error in the event locations, and the size of the search bin area. The significances obtained then reflect the balance between the clustering of the signal at the scales of the search bin radius, and the scales at which random accumulation of background events is more probable. A discovery (i.e. a p-value $< 2.8 \times 10^{-7}$ ) is achieved when sampling the region with bins of $1.5^{\circ}$ radius (figure \ref{fig:2c}), but the significance drops considerably when observing at different angular scales, and any information regarding the spatial structure of the signal within the region under examination is lost. However, this information is recovered when we study the spatial correlations between events with MPS. The results of the analysis of the region with MPS are shown in figure \ref{fig:3}. Figure \ref{fig:3a} shows the clustering function of the events up to $5^{\circ}$ scales, and figure \ref{fig:3b} shows the significances obtained at each angular scale. The measured spatial correlations between events below $5^{\circ}$ follow a power-law with exponential decay, consistent with the signal field that was used in the simulations, and the discovery of the simulated astrophysical neutrino signal goes up to angular scales of $\sim 2.5^{\circ}$. Therefore, MPS results have the potential to discover a more complicated structure of a possible astrophysical neutrino signal than what was observed with the standard scan of the region.

\subsection{Low S/N cases}
\label{ss:5.2}

In this example a total of 40 signal events were injected in the region. With a background expectation of 207 events within R, this situation corresponds to a background dominated sample. This case exemplifies a more realistic situation, since a total of 40 signal events present at the final sample of neutrino telescopes from a region of $10^{\circ} \times 10^{\circ}$ constitutes a more realistic situation given the current effective areas of neutrino detectors and sources with moderate neutrino fluxes \cite{IceCube40}. 

\begin{figure*}[!ht]
\centering
\subfigure[Significance map of event densities at $1^{\circ}$ scales in sample 1]{\label{fig:4a}\includegraphics[width=0.3\textwidth,keepaspectratio]{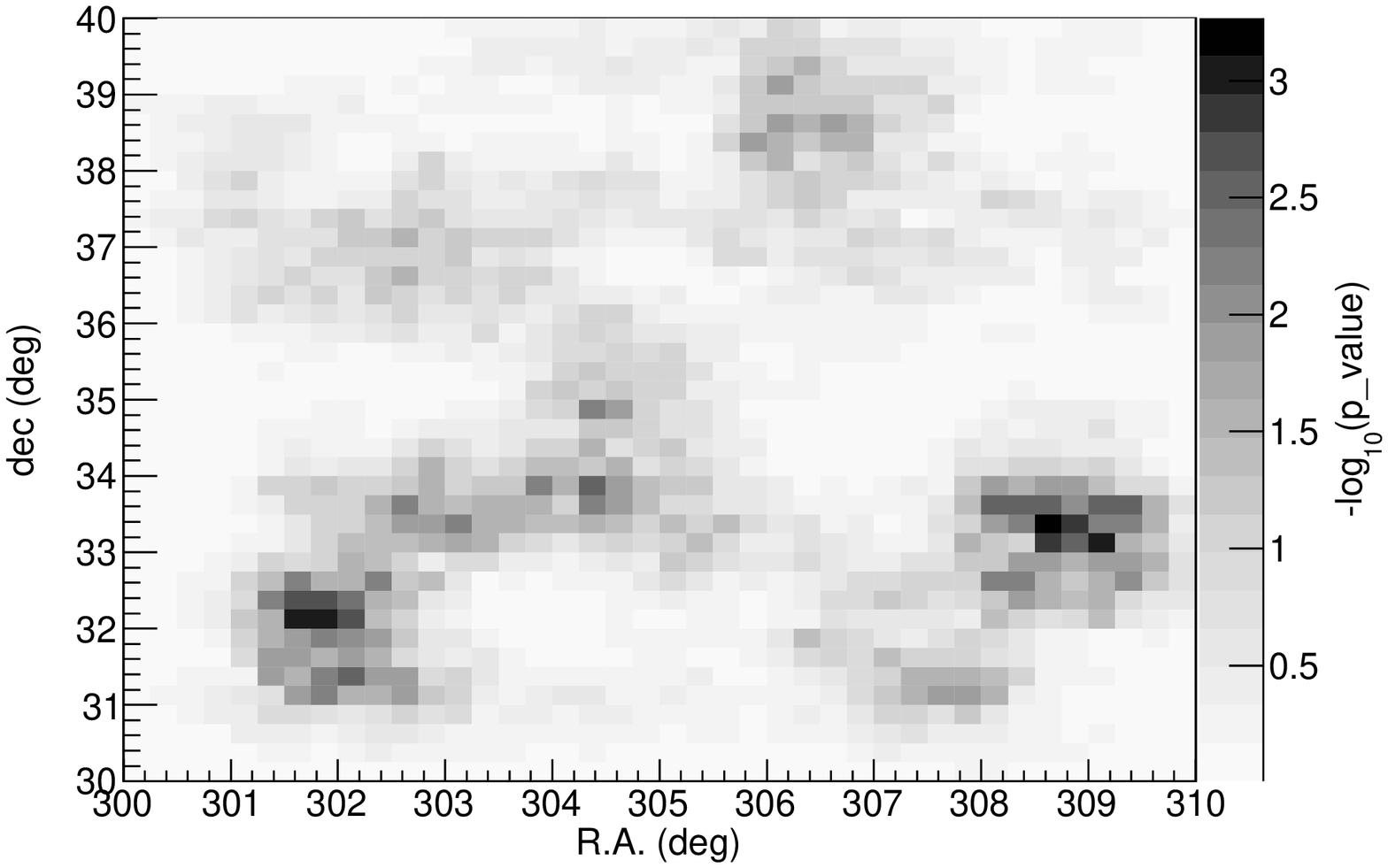}} \hspace{20pt} 
\subfigure[Significance map of event densities at $1^{\circ}$ scales in sample 2]{\label{fig:4b}\includegraphics[width=0.3\textwidth,keepaspectratio]{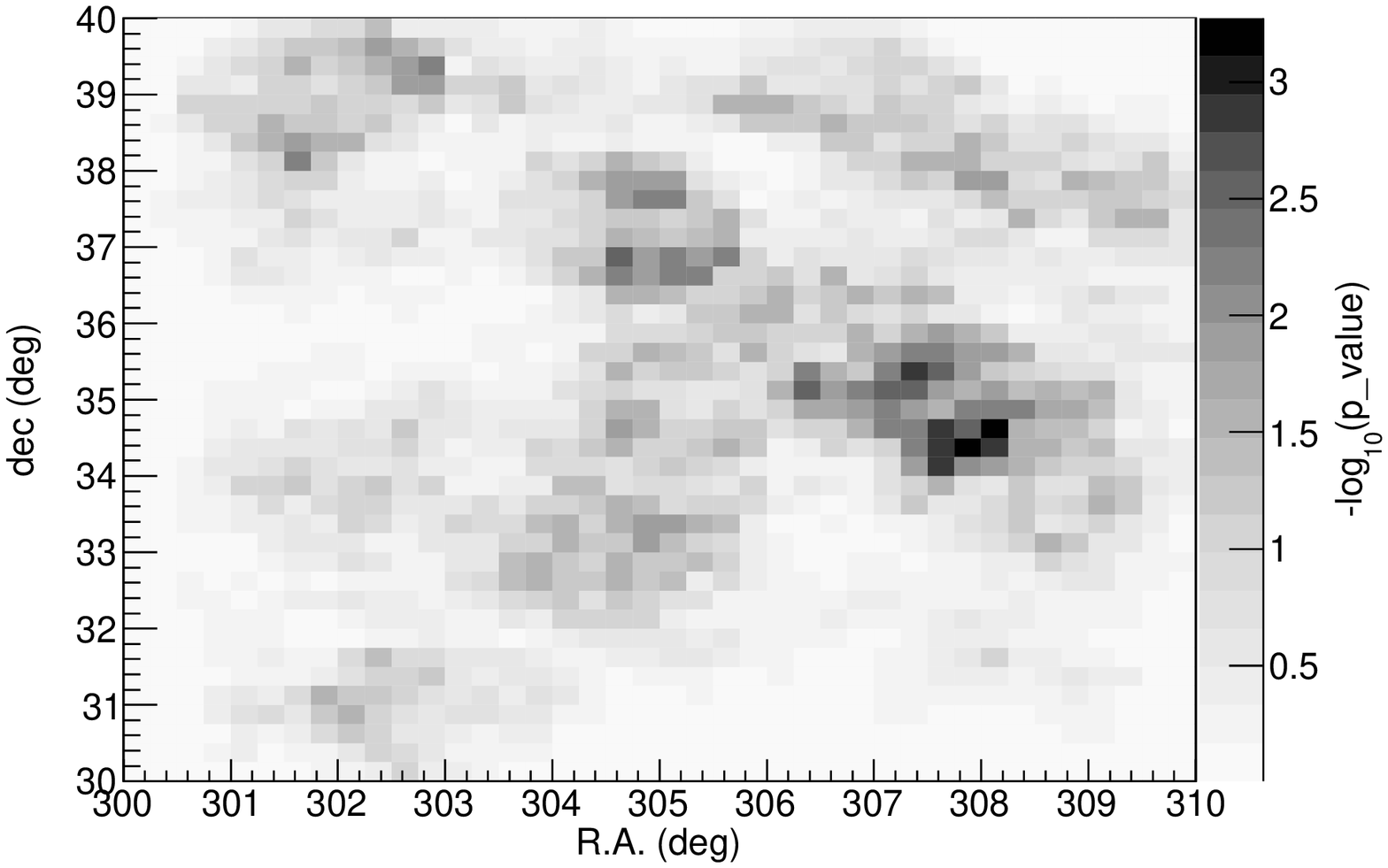}} \hspace{20pt}
\subfigure[Significance map of event densities at $1^{\circ}$ in sample 3]{\label{fig:4c}\includegraphics[width=0.3\textwidth,keepaspectratio]{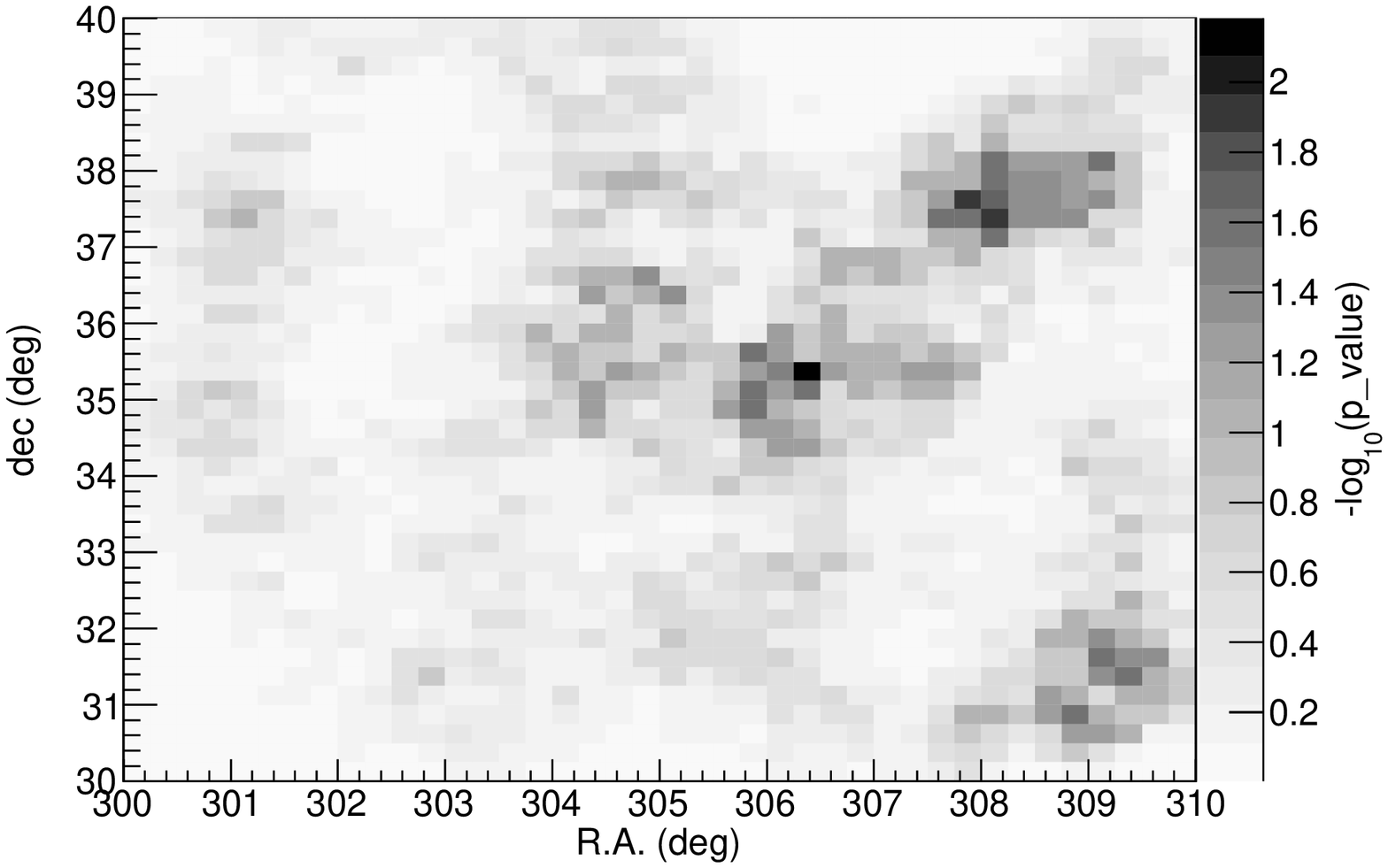}} \hfill

\caption{Results from the binned scan over signal+background samples}
\label{fig:4}
\end{figure*}

\begin{figure*}[!ht]
\centering
\subfigure[Significance map of event densities at $1^{\circ}$ scales in sample 1]{\label{fig:5a}\includegraphics[width=0.3\textwidth,keepaspectratio]{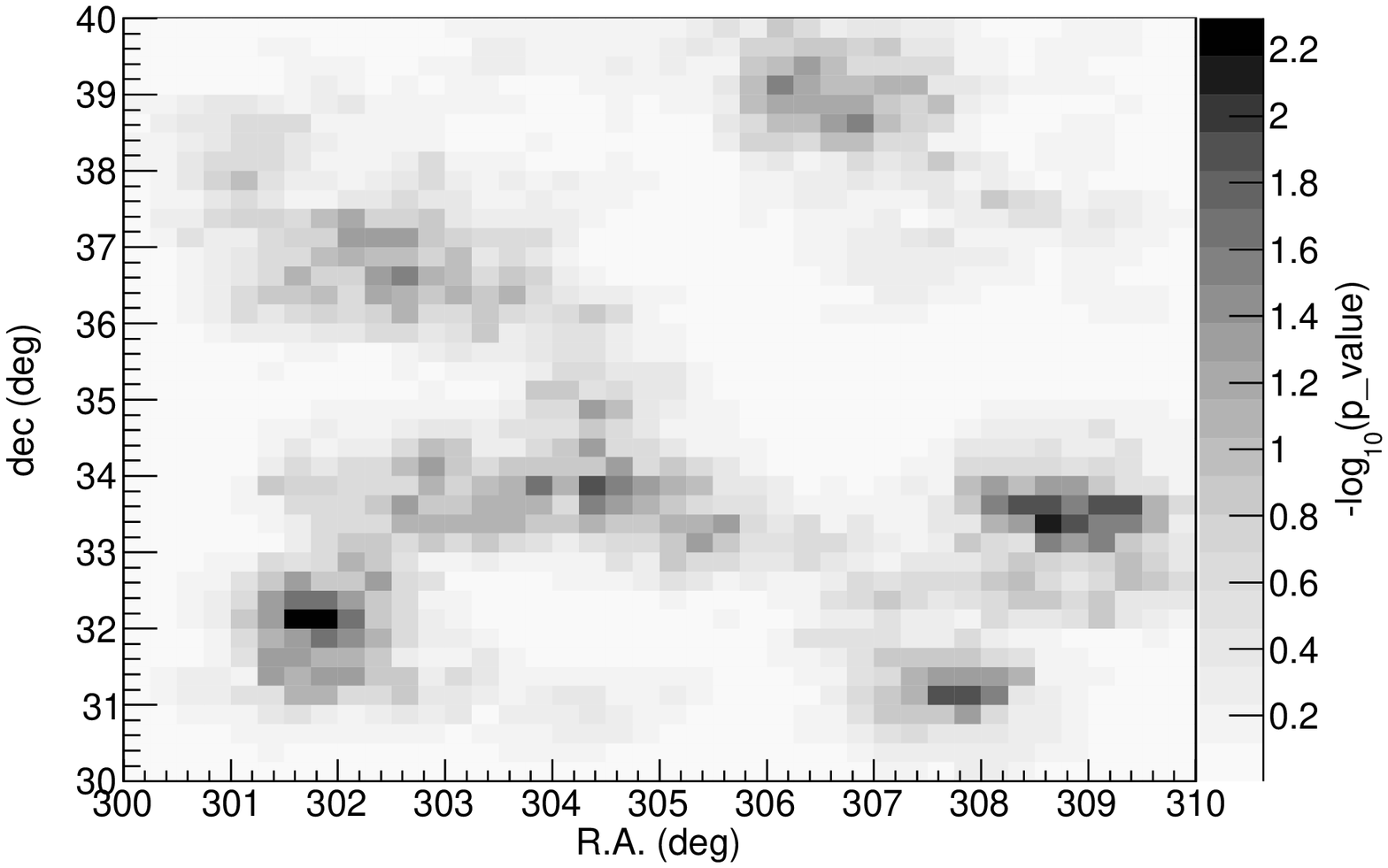}} \hspace{20pt} 
\subfigure[Significance map of event densities at $1^{\circ}$ scales in sample 2]{\label{fig:5b}\includegraphics[width=0.3\textwidth,keepaspectratio]{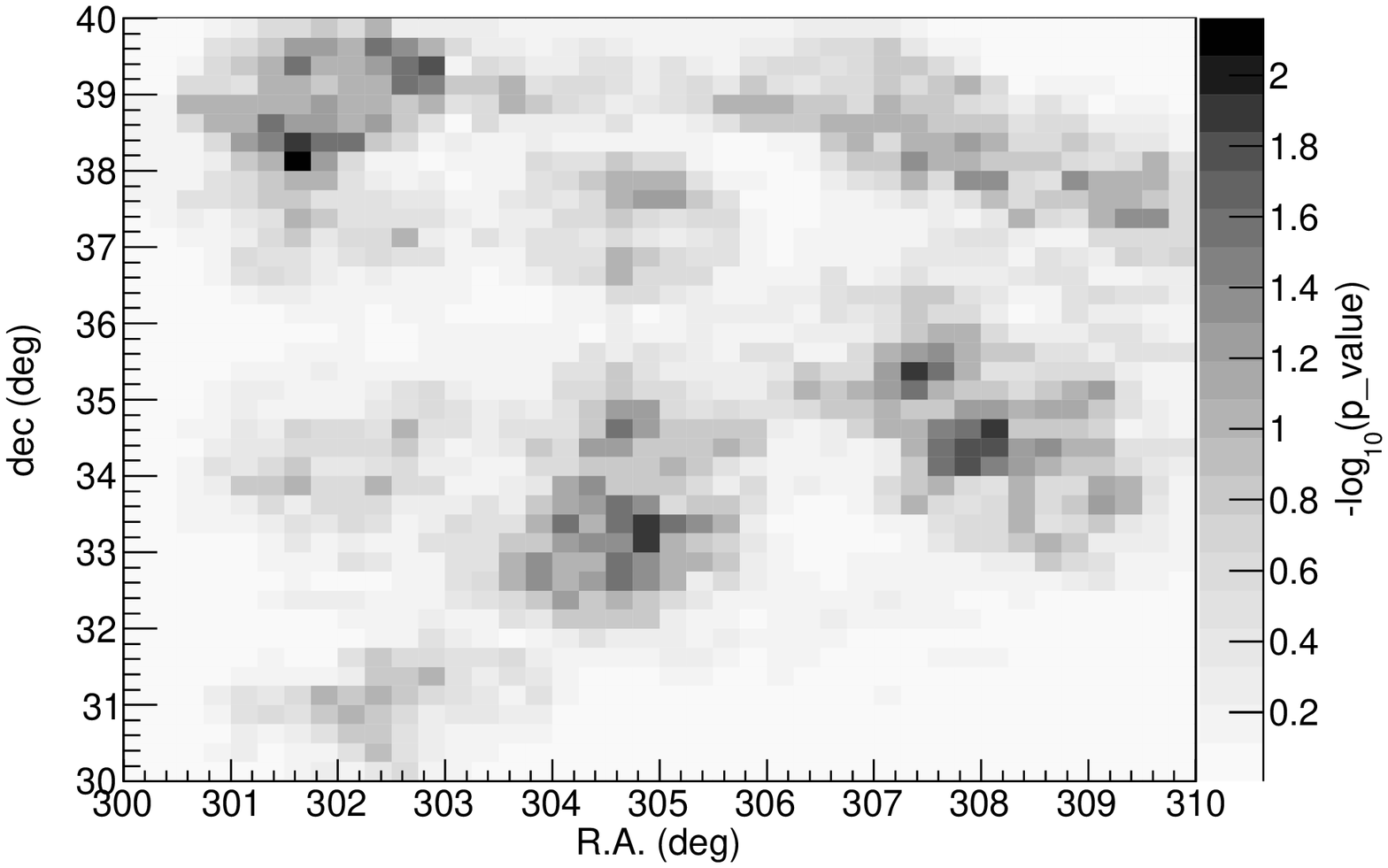}} \hspace{20pt}
\subfigure[Significance map of event densities at $1^{\circ}$ in sample 3]{\label{fig:5c}\includegraphics[width=0.3\textwidth,keepaspectratio]{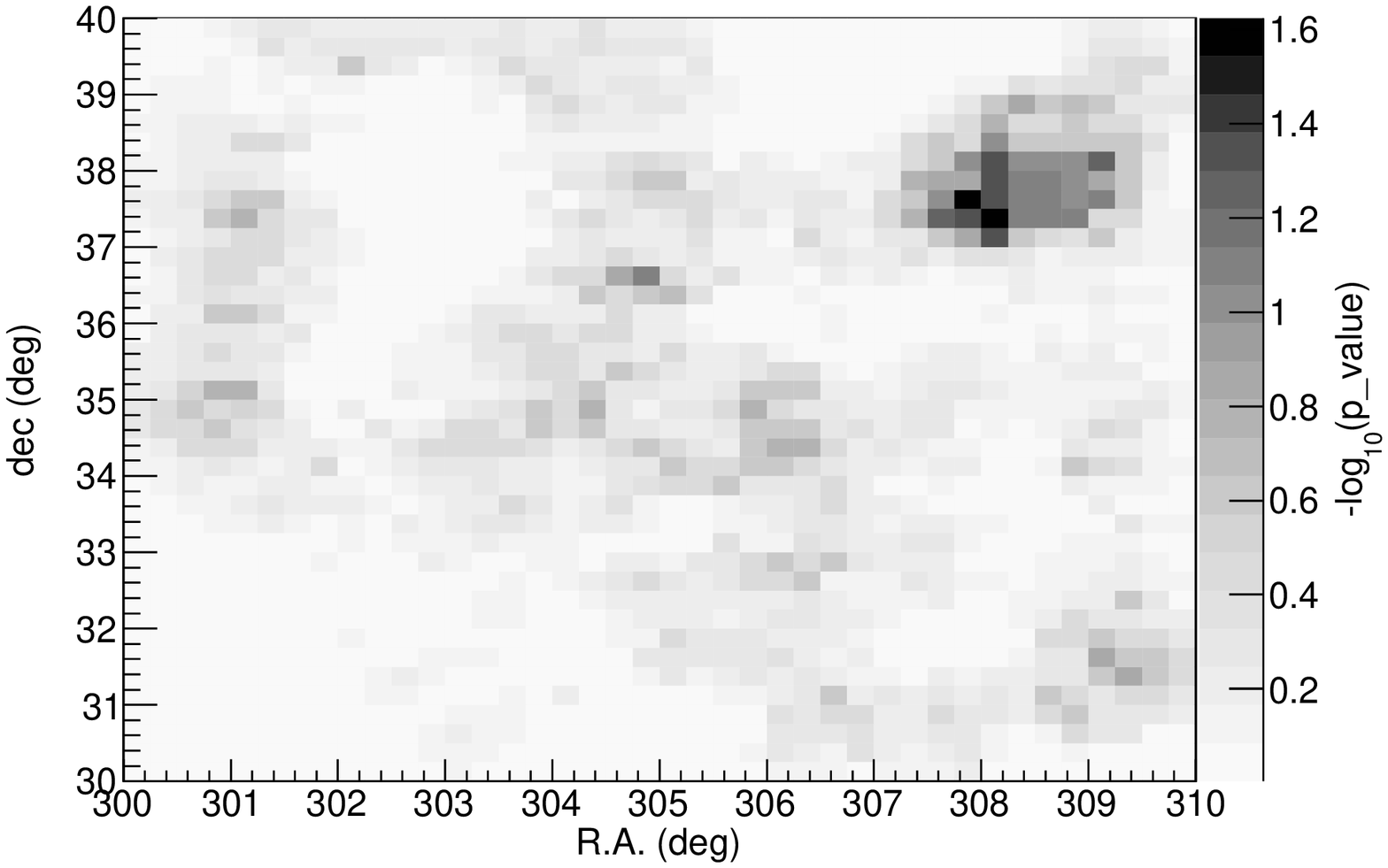}} \hfill

\caption{Results from the binned scan over background-only samples}
\label{fig:5}
\end{figure*}

\begin{figure*}[!ht]
\centering
\subfigure[Significance map of event densities at $0.5^{\circ}$ scales in the combined sample]{\label{fig:6a}\includegraphics[width=0.3\textwidth,keepaspectratio]{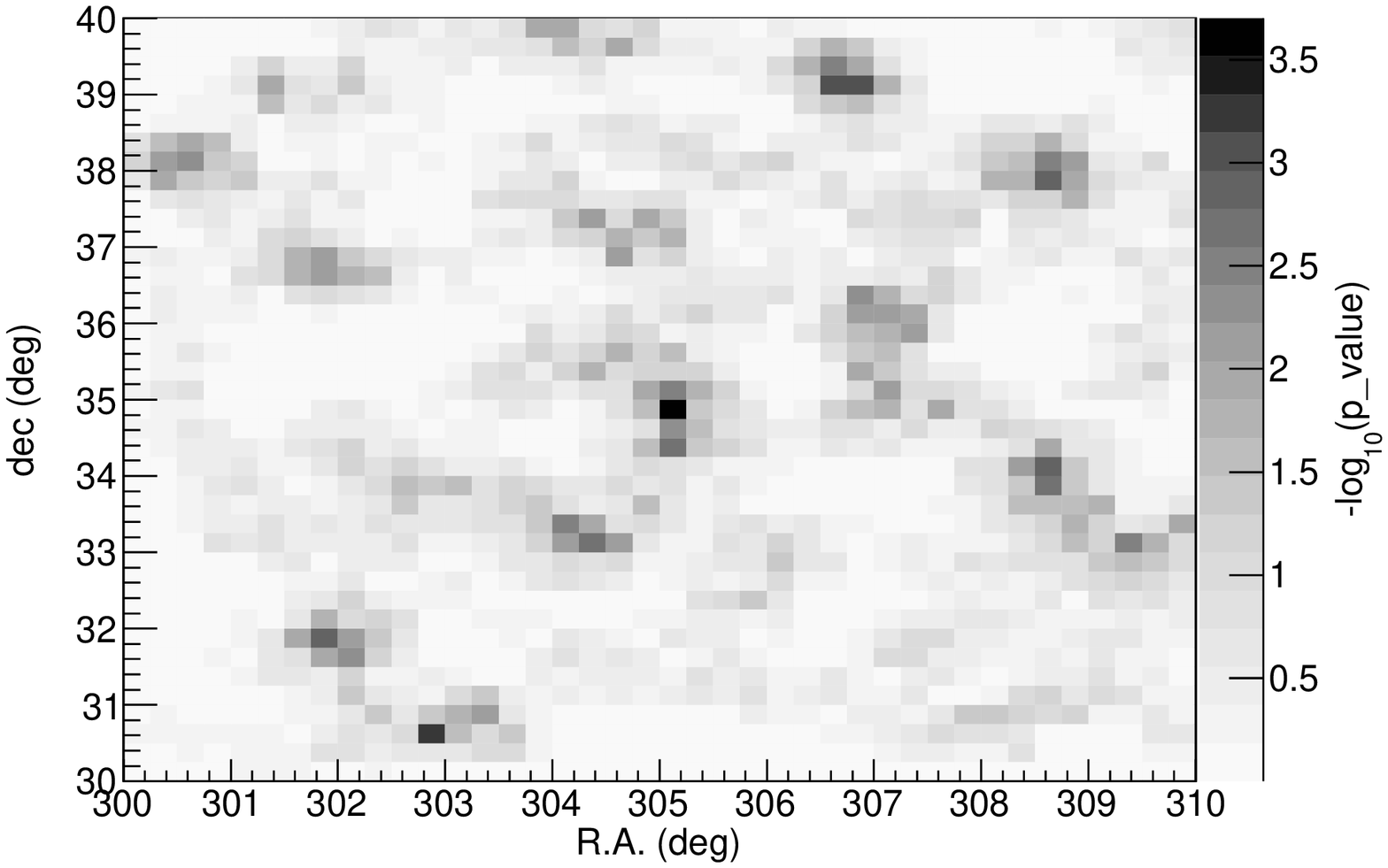}} \hspace{20pt}
\subfigure[Significance map of event densities at $1^{\circ}$ scales in the combined sample]{\label{fig:6b}\includegraphics[width=0.3\textwidth,keepaspectratio]{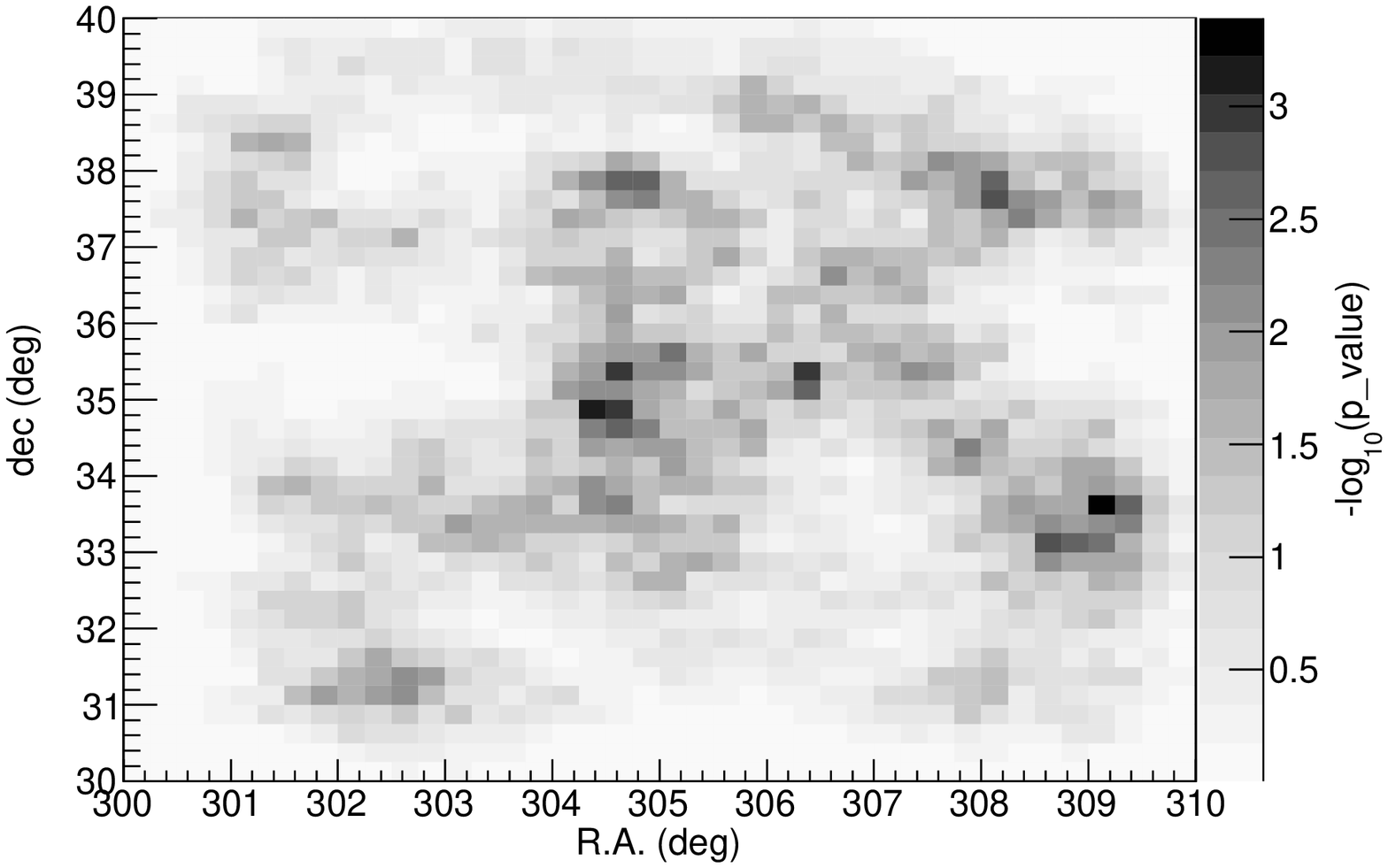}} \hspace{20pt}
\subfigure[Significance of the clustering function at each of the angular scales tested in the combined sample.]{\label{fig:6c}\includegraphics[width=0.3\textwidth,keepaspectratio]{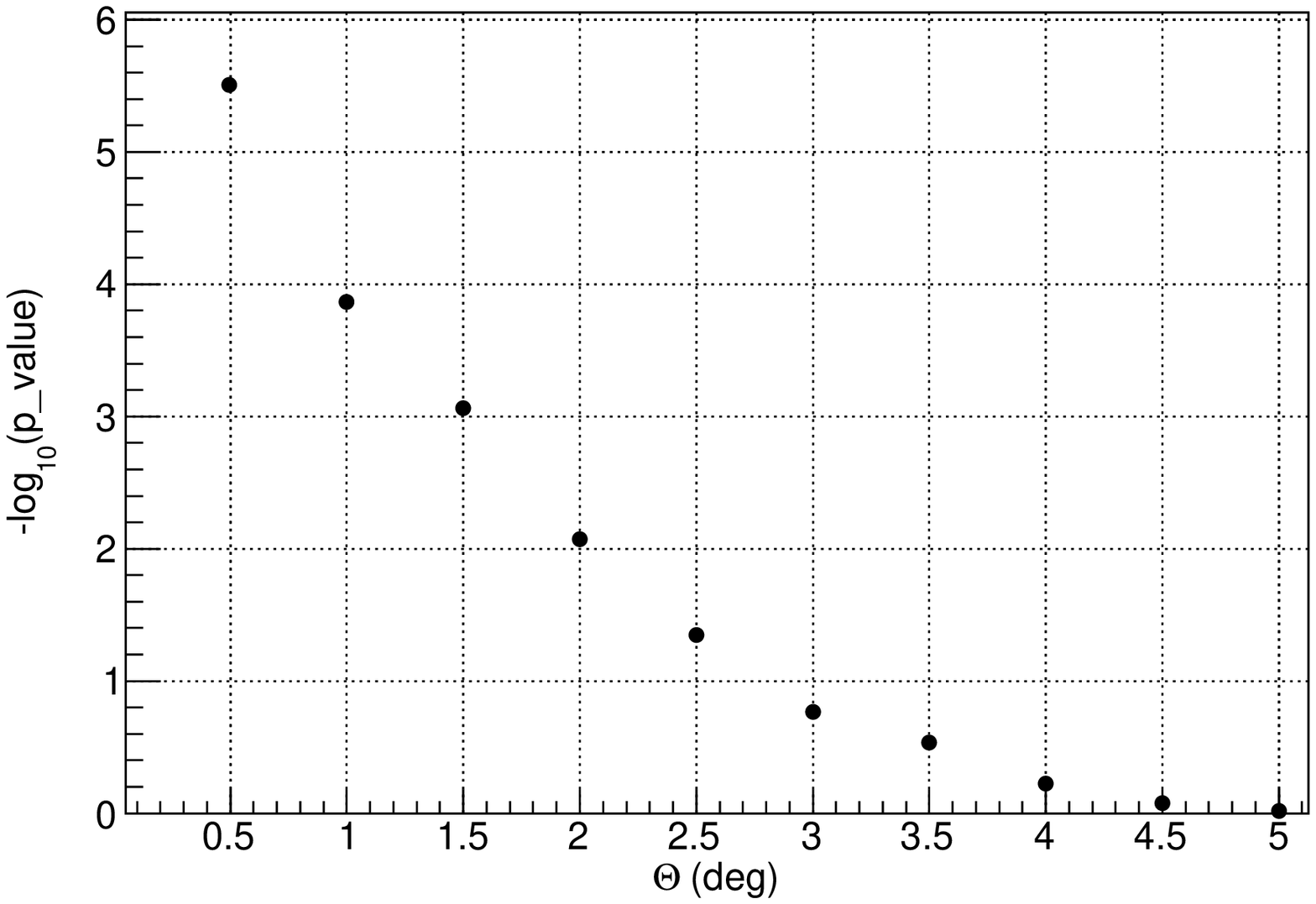}} \hfill \\ 
\caption{Results from the binned scan and from MPS over the combined sample}
\label{fig:6}
\end{figure*}

We illustrate here three simulated examples of such situation. Fig.~\ref{fig:4} shows the result of a $1^{\circ}$ binned scan for three realizations of the background field plus 40 signal events following the spatial distribution of figure \ref{fig:1}. The three cases are compatible with background fluctuations, reaching a 3$\sigma$ level at most. Fig.~\ref{fig:5} shows the same scan, but in which signal events have been removed. The comparison of figs.~\ref{fig:4} and \ref{fig:5} manifests the random appearance of events even in the signal+background maps, where the most significant spots occurred always at random locations, as expected in background-only skymaps, but with a slightly higher significance due to the additional presence of signal events in the region. If we combine the three simulated samples (figures \ref{fig:6a},~\ref{fig:6b}), corresponding to a larger period of data taking in the real situation, the S/N is still too low to produce a significant detection from the binned scan, and the minimum p-value achieved only reaches 3$\sigma$ at $0.5^{\circ}$ scales (figure \ref{fig:6a}). If, in addition, we correct from the trials associated to the scan of an extended region, this significance is further reduced. However, although the significance of these individual spots is not enough for a detection, their distribution is not common of a random distribution of background events. This is illustrated when we use MPS to quantify the departures of the whole event pattern within R from a distribution of background events. With MPS, the minimum p-value of the observed event pattern reaches $4.3\sigma$ observation (figure \ref{fig:6c}). The trial factors from scanning several clustering scales with MPS are very small, and the MPS post-trial significance remains practically unchanged, from the pre-trial p-value of $10^{-5.5}$ to a post-trial p-value of $10^{-5.3},$ yielding also a $4.3\sigma$ final result. 

Therefore, in the typical situations of neutrino telescopes of low S/N, the only hint of a possible neutrino signal would be the accumulation of relatively weak spots within the region which is believed to be a factory of cosmic rays. As long as the S/N is low, these most significant spots do not necessarily have to appear at the same location in the analysis of the same region in different epochs, since they have an important background component, which fluctuates randomly within the region. The signal would be missed by standard scan analysis, but could be detected by correlation analysis like MPS. Eventually, after some years of integration time, the S/N would be high enough to overcome the background fluctuations, providing an image of the neutrino emission in the region. But before this situation is achieved, we expect to find a distribution of neutrino events which departures significantly from the average distribution of background events, provided there is some clustering of neutrino sources within the region under examination.

\section{Discussion}
\label{s:5}

The discovery of high energy neutrino emission from a particular region of the Galaxy would suggest a local injection of high energy protons or heavier nuclei; as well as their interaction not very far away from their sources of origin, before they lose their energy by diffusion in Galactic magnetic fields \cite{Torres2011, Ahlers09}. Whereas the discovery of individual point sources of high energy neutrinos seems challenging because of the limited sensitivity of neutrino detectors, the agglomeration of neutrino sources inside the same region of the sky could be detected as a significant departure from the distribution of atmospheric neutrino events. Such clustering of neutrino sources would be a natural consequence of the distribution of stellar matter in the Galaxy. Massive and young stellar populations, the birth sites of Galactic core-collapse supernova, are associated to the spiral arms of the Milky Way, following a hierarchical scheme where stars are grouped in clusters, and the stellar clusters themselves form part of open cluster complexes. 

Under the assumption that the explosion of a supernova is the most plausible energy source for cosmic ray acceleration in the Galaxy, we propose in this paper to search for neutrino emission in the regions where massive stars die, that is, where the cosmic ray acceleration process takes place in the supernova remnant shocks that form around the explosion site. Given the distribution of massive star clusters in the Galaxy, we can find such potential cosmic ray factories at angular scales which could even reach around $\sim 10^{\circ}$ or more, depending if we also want to consider clustering of potential accelerators due to projection effects. This type of search therefore implies the analysis of extended regions, much larger than the detector's angular resolution. We emphasize the usefulness of a search method which accommodates different degrees of complexity in the spatial distribution of astrophysical neutrino events, and try to not restrict ourselves only to the case of point-like emission. In this paper, we propose a method to search for neutrino emission from extended regions which takes into account all these considerations, and that we conveniently named the Multi Point Source (MPS). Comparing to a single source search, the MPS would give a better result only if the emission deviates significantly from spherical symmetry, whereas it would be comparable to a standard binned search in the opposite case.

In situations in which the search for extraterrestrial neutrinos goes beyond the single and spherically symmetric source approach, we have shown that correlation analysis like MPS are able to discover a significant neutrino event pattern inside the region under investigation with a higher significance than standard searches, which scan the region looking for single sources. 

In simulated examples of a high signal to noise ratio, where also a standard scan of the region results in a discovery, the MPS is proven to be worth performing in order to extract information of the neutrino event pattern at different angular scales, where no significant information is extracted from the scan. Even more interesting, given the current status of neutrino astronomy, is the case of a very low S/N, where we have shown that even when the standard scan of a particular patch of the sky do not yield any hint of an astrophysical signal, MPS indicates the existence of astrophysical neutrino signal inside the region under study. 

The conclusions of this paper about the potential of the search method we propose to discover neutrino emission with respect to standard search strategies are independent on the spatial correlation function that we adopted for the simulations. The examples used here are hypothetical, and not attempt to describe the neutrino emission from any particular region of the Galaxy, but to illustrate the usefulness of our method for detecting a significant event pattern even in strongly background dominated samples. \\ \\

The authors are thankful to Aday Robaina for useful discussions about the two-point correlation function. Yolanda Sestayo acknowledges the financial support from J.M. Paredes through ICREA Academia.

\end{document}